\documentstyle[twoside,fleqn,espcrc2,epsfig]{article}

\newcommand{\bea}{\begin{eqnarray}}
\newcommand{\eea}{\end{eqnarray}}
\def \simlt {\mathrel{\vcenter{\hbox{$<$}\nointerlineskip\hbox{$\sim$}}}}

\title{Higgs pair production in the MSSM with explicit CP violation}

\author{D. A. Demir \address{The Abdus Salam International Center for Theoretical Physics, I-34100,
Trieste, Italy}} 

\begin{document}

\begin{abstract}
In the minimal supersymmetric standard model with explicit CP violation, associated
production of the lightest Higgs boson with heavier ones is analyzed. Due to explicit 
CP violation, the Higgs bosons are no longer CP eigenstates so that both of the heavy Higgs 
bosons contribute to the process. While the radiative corrections in the Higgs sector
turn out to be quite important, the vertex radiative corrections remain small as in the 
CP conserving theory.

\end{abstract}
\maketitle
\section{Introduction}
The Lagrangian of the minimal supersymmetric standard model (MSSM) consists of
various soft masses as well as the $\mu$ parameter coming from the superpotental
\cite{rosiek}. In principle, gaugino masses, trilinear couplings, Higgs bilinear 
coupling and the $\mu$ parameter can be complex; however,  
not all these phases are physical \cite{phase1}.  Indeed, with the global symmetries 
of the MSSM Lagrangian, the physical phases could be reduced to those of 
the CKM matrix, trilinear couplings $A_{f}$, and the $\mu$ parameter \cite{phase1,phase2}. 

Recently, the supersymmetric CP phases $\gamma_{\mu}\equiv {\cal{A}}rg\{\mu\}$ 
and $\gamma_{A_{f}}\equiv{\cal{A}}rg\{A_{f}\}$ have gotten much 
interest in Higgs phenomenology \cite{ben1,ben2,wagner}, 
FCNC processes \cite{biz-korean}, and electroweak baryogenesis \cite{quiros}. 
Here we discuss the effects of these supersymmetric CP phases on the associated Higgs production 
by studying $e^{+}e^{-}\rightarrow H_{CP\; even}H_{CP\; odd}$ scattering. 
This process was already discussed in \cite{pokorski} in the CP--conserving MSSM
with one--loop accuracy, and the radiative corrections were found be
at most $\sim 5\%$. We analyze the same process in the CP--violating MSSM by 
including the radiative corrections in the Higgs sector as well as 
$Z  H_{CP\; even}H_{CP\; odd}$ vertex, where photon contribution is neglected as it is much smaller.
Results agree with \cite{pokorski} in that the vertex corrections are quite small; 
however, radiative corrections in the Higgs sector are important. Below, 
in accordance with EDM \cite{phase1,olive1} and cosmological constraints \cite{olive2}, 
$\gamma_{\mu}$ will be set to zero in course of the analysis whereas 
$\gamma_{A}\equiv \gamma_{A_{t}}$ will taken as free.

\begin{figure}[htd]
\begin{center}
\vspace{9pt}
\epsfig{file=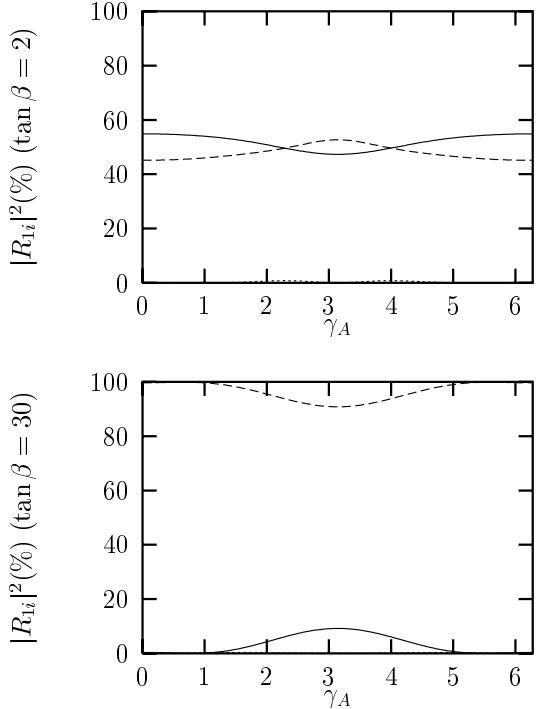,width= \linewidth}
\caption{$|R_{11}|^{2}$ (solid), $|R_{12}|^{2}$ (dashed) and $|R_{13}|^{2}$ (dotted) 
in percents for $\tan\beta=2$ (up) and $\tan\beta=30$ (down).}
\end{center}
\end{figure}

\begin{figure}[htd]
\begin{center}
\vspace{9pt}
\epsfig{file=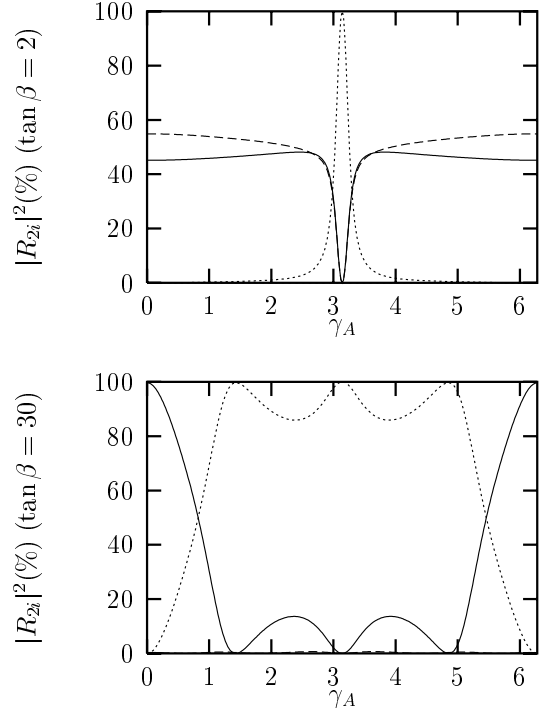,width= \linewidth}
\caption{The same as Fig. 1 but for $H_{2}$.}
\end{center}
\end{figure}

\section{$e^{+}e^{-}\rightarrow H_{CP\; even}H_{CP\; odd}$ Scattering}
The supersymmetric CP phases $\gamma_{A}$ and $\gamma_{\mu}$ show up in sfermion, 
chargino and neutralino mass matrices so that the processes involving these
particles as well as their loop effects depend on these phases explicitly 
\cite{biz-korean,ben1,ben2,wagner}. 

\begin{figure}[htd]
\begin{center}
\vspace{9pt}
\epsfig{file=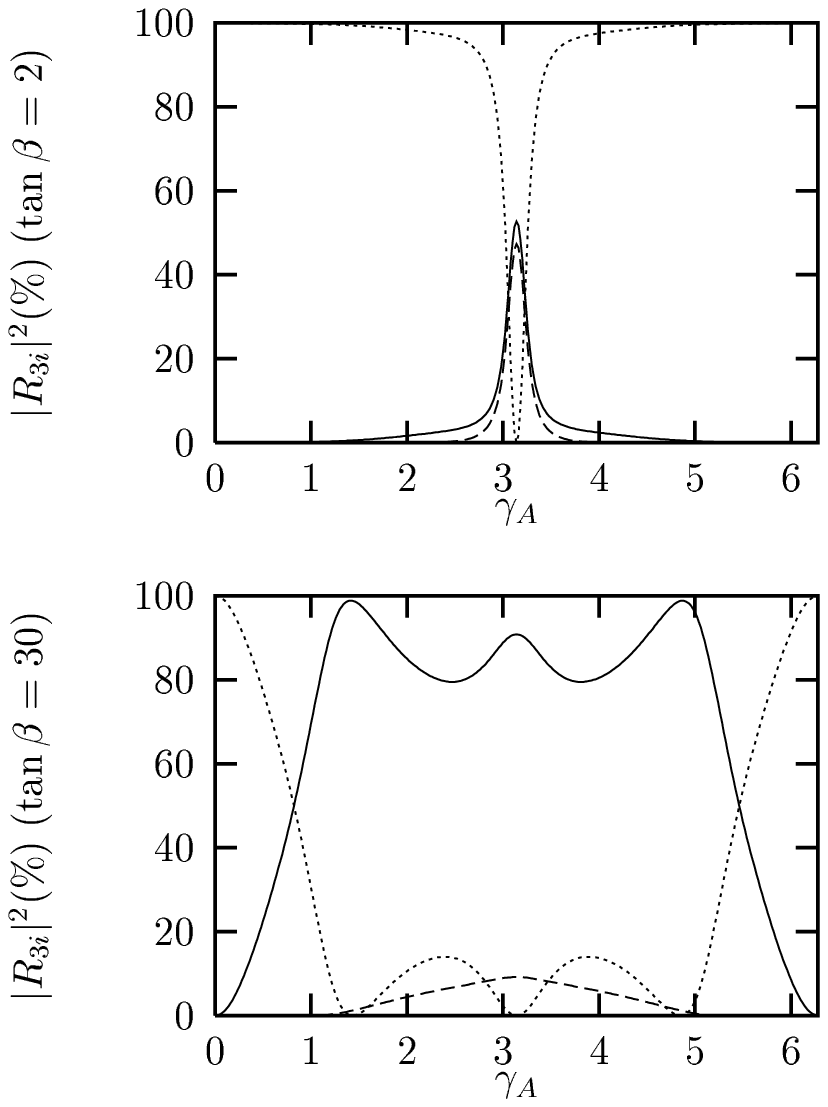,width= \linewidth}
\caption{The same as Fig. 1 but for $H_{3}$.}
\end{center}  
\end{figure}

\begin{figure}[htd]
\begin{center}
\vspace{9pt}
\epsfig{file=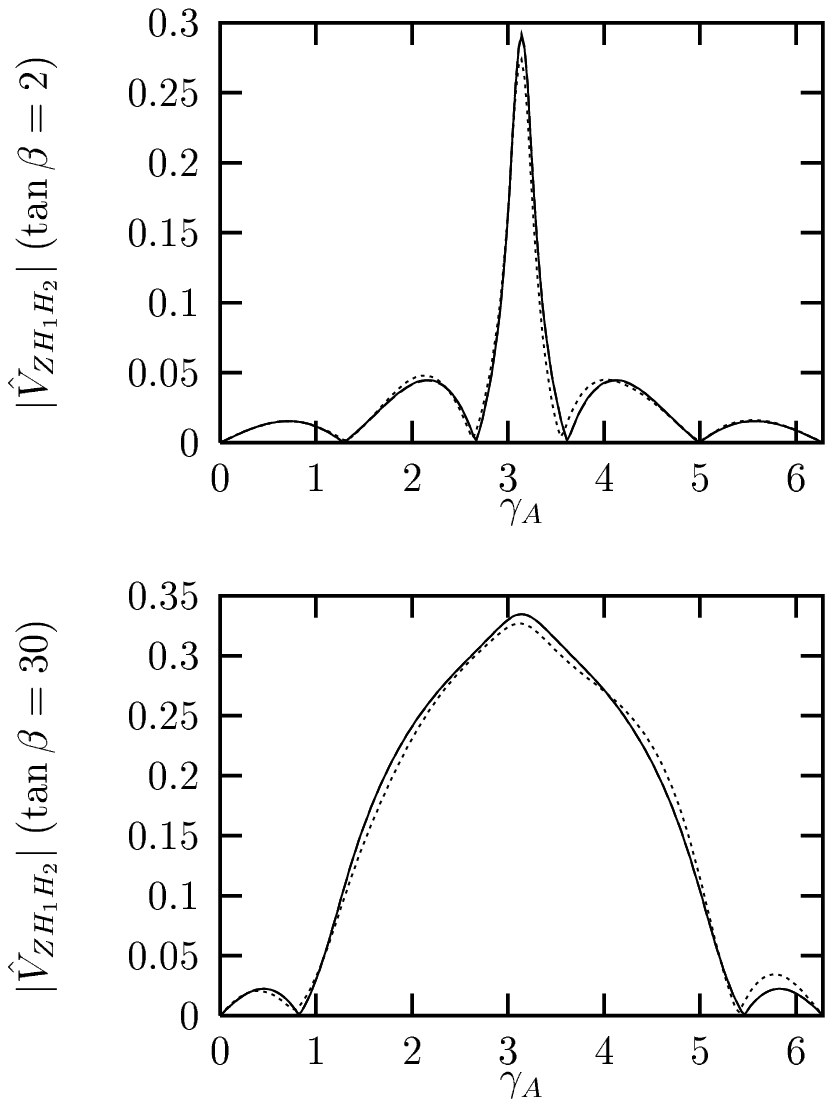 ,width= \linewidth}
\caption{Variation of $Z H_{1} H_{2}$ coupling with $\gamma_{A}$ when
there is no vertex correction (solid), with only top quark contribution (dashed),
and with full vertex correction (dotted).}
\end{center}
\end{figure}

The loop effects of the MSSM particles on the Higgs potential could be parametrized 
in a simple and elegant way by using the effective potential approximation \cite{effpot}.
For $\tan\beta\simlt 40$, the radiative corrections to the potential is dominated by 
the top quark and top squark loops as discussed in \cite{ben1,ben2}. Diagonalization of the
top squark matrix is accomplished by a unitary transformation with a unitary matrix
$U_{\tilde{t}}$ which can be parametrized by the top squark left--right mixing angle
$\theta_{\tilde{t}}$ and the phase $\gamma_{t}\equiv {\cal{A}}rg\{A_{t}-\mu \cot\beta\}$ \cite{ben1,ben2}. 

\begin{figure}[htd]
\begin{center}
\vspace{9pt}  
\epsfig{file=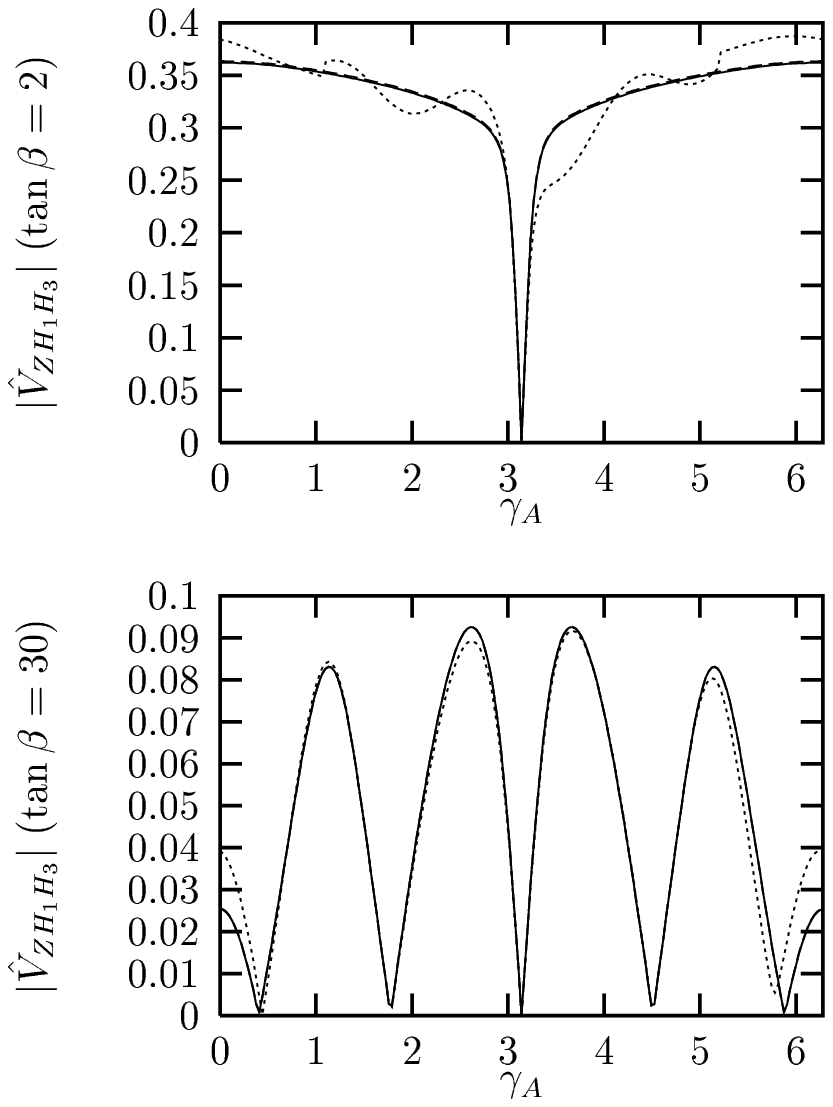 ,width= \linewidth}
\caption{The same as Fig. 4 but for $Z H_{1} H_{3}$ coupling.}
\label{spect8}
\end{center}  
\end{figure}  

On the other hand, the mass--squared matrix of the Higgs scalars, after including the one--loop radiative corrections, is a
$3\times 3$ matrix (in the basis ${\cal{B}}=\{\phi_{1}\equiv {\cal{R}}e[H_{1}^{0}], \phi_{2}\equiv {\cal{R}}e[H_{2}^{0}], A\equiv
\sin\beta {\cal{I}}m[H_{1}^{0}] + \cos\beta {\cal{I}}m[H_{2}^{0}]\}$) having all entries nonvanishing. In particular, those
entries of the matrix connecting $\phi_{1,2}$ to $A$ are proportional to $\sin (\gamma_{A}+\gamma_{\mu})$\footnote{Here we
neglect the radiatively--induced alignment between the Higgs doublets \cite{benim}.}, and they break CP
invariance in the
Higgs sector explicitly \cite{ben2}. Diagonalization of the Higgs mass--squared matrix can be accomplished with a $3\times 3$
orthonormal matrix $R$ as $(H_{1}, H_{2}, H_{3})=R^{T}(\phi_{1},\phi_{2}, A)$. Here $H_{i=1,2,3}$ refer to the mass eigenstate
Higgs scalars, and each of them has odd-- and even--CP compositions. Therefore, none of the mass eigenstate Higgs bosons is a CP
eigenstate. 

In illustrating the CP compositions of the mass eigenstate Higgs particles, and analyzing the $Z H_{CP\; even} H_{CP\; odd}$ 
vertex  $\gamma_{A}$ and $\tan\beta$ will be left free; however, the remaining parameters are fixed as: $
M_{\tilde{L}}=M_{\tilde{R}}=500\; \mbox{GeV};\, |A_{t}|= 1.3\; \mbox{TeV};\,|\mu|=\mu=250\; \mbox{GeV};\,\tilde{M}_{A}=200
\mbox{GeV}$ \cite{ben1}. In presenting the figures our convention will be such that $H_{1}$, $H_{2}$ and $H_{3}$ reduce,
respectively, to the lightest CP--even, heavy CP--even, and pseudoscalar Higgs particles when $\gamma_{A}\rightarrow 0$. 

Fig. 1 shows the CP compositions of the lightest Higgs $H_{1}$ as a function of $\gamma_{A}$ for $\tan\beta=2$ (upper window)
and $\tan\beta=30$ (lower window). As the figure suggests, for $\tan\beta=2$, CP--even components of $H_{1}$ oscillate between $\sim
45\%$ and $\sim 55\%$ in the entire $\gamma_{A}$ range while its CP--odd component remains below $1\%$ everywhere. On the other
hand, for $\tan\beta=30$, $|R_{12}|^{2}$ remains above $\sim 90\%$, and correspondingly $|R_{11}|^{2}$ is always below $\sim
10\%$. This time its CP--odd component is even smaller; it never exceeds  $\sim 0.5\%$. This rearrangement of the various CP
compositions results from the increase in $\tan\beta$ which decouples light and heavy sectors. 
 
Depicted in Fig. 2 are the CP compositions of $H_{2}$. Away from $\gamma_{A}=\pi$ behaviour of all three curves follows 
from Fig. 1. For $\gamma_{A}\sim \pi$, however, $H_{2}$ becomes a pure CP--odd particle contrary to its pure CP--even character
at $\gamma_{A}=0$. This change in the CP--purity of $H_{2}$ as $\gamma_{A}$ switches from one CP--conserving point to the next
results from change in the strength of the radiative corrections; for the same parameter set the light top squark is much
lighter at $\gamma_{A}=\pi$ than $\gamma_{A}=0$ \cite{ben1}. Fig. 3 shows the CP compositions of $H_{3}$, and its interpretation
is similar to that of $H_{2}$ since these two have opposite CP quantum numbers at the CP--conserving points. 

After specifying the CP compositions of the mass eigenstate Higgs bosons $H_{i}$ in Figs. 1-3, we now turn to the discusion of
$e^{+}e^{-}\rightarrow H_{CP\; even}H_{CP\; odd}$ scattering by studying $Z H_{CP\; even}H_{CP\; odd}$ vertex. In
particular, we discuss the production of $H_{2}$ and $H_{3}$ in association with $H_{1}$ to observe the effects of odd-- and
even--CP compositions. It is possible to realize $H_{2}H_{3}$ production as well; however, for a
$\sqrt{s}=500\, \mbox{GeV}$ $e^{+}e^{-}$ collider, as we consider here, this mode is phase space suppressed. The radiative
corrections in the Higgs sector is not the whole story because one has to include vertex radiative corrections too
\cite{pokorski}. In this respect, consistent with the approximations for the Higgs sector, we consider top quark and top squark
triangles in computing the one--loop vertex corrections \cite{ben1}. Contrary to the CP--conserving theory \cite{pokorski}, here
the explicit CP--violation allows for a pseudoscalar to couple to identical top squarks, which, when the latter is light,
enhances the vertex radiative corrections \cite{ben1}.

Fig. 4 illustrates the absolute magnitude of $Z H_{1}H_{2}$ coupling $|\hat{V}_{H_{1}H_{2} Z}|$ when there is no vertex
correction (solid curve), when
only the top quark loop is considered (dashed curve), and when both top quark and top squark loops are included (dotted curve).
For $\gamma_{A}=0$ this vertex necessarily vanishes as both Higgs bosons are CP--even. As $\gamma_{A}$ increases, however,this
vertex starts having non--vanishing values due to the CP--compositions of $H_{1}$ and $H_{2}$ in Figs. 1 and 2. A glance at the
upper windows of Figs. 1, 2 and 4 shows that, even if the CP--odd compositions are much smaller than the CP--even ones, since 
$|\hat{V}_{H_{1}H_{2} Z}|$ vertex involves the multiplication of these two, one has an enhancement in the coupling. This
statement
remains valid also for $\tan\beta=30$, as can be seen from the lower windows of the same figures. In both windows of Fig. 4 one
observes the maximization of $|\hat{V}_{H_{1}H_{2} Z}|$ at $\gamma_{A}=\pi$ which, obviously, follows from the opposite CP
purities of $H_{1}$ and $H_{2}$ at this point. One notices that, for both values of $\tan\beta$, vertex radiative corrections 
are small compared to raditive corrections in the Higgs sector. This has different reasons for different $\tan\beta$ values. For
$\tan\beta=2$, away from $\gamma_{A}=\pi$, CP--odd compostions of $H_{1}$ and $H_{3}$ are small, and this suppresses the vertex
radiative corrections. For $\gamma_{A}\leadsto \pi$, radiative corrections are given by the ones in the CP--conserving theory
\cite{pokorski}, which are already small. For $\tan\beta=30$, despite rather enhanced CP purities of $H_{1}$ and $H_{2}$,
$|\hat{V}_{H_{1}H_{2} Z}|$ remains at similar magnitude with its value for $\tan\beta=2$. This is due to the suppression of
$\phi_{2}$ component in  $|\hat{V}_{H_{1}H_{2} Z}|$  with large $\tan\beta$ as it is proportional to $\cos\beta$ \cite{ben1}.

Depicted in Fig. 5 is the absolute magnitude of $Z H_{1}H_{3}$ coupling $|\hat{V}_{H_{1}H_{3} Z}|$ when there is no vertex
correction (solid curve), when
only the top quark loop is considered (dashed curve), and when both top quark and top squark loops are included (dotted curve).
This vertex remains necessarily non--vanishing as $\gamma_{A}\leadsto 0$ since $H_{1}$ and $H_{3}$ become, respectively, the
lightest and th epseudoscalar Higgs bosons of the CP--conserving theory. On the other hand, as is obvious from Fig. 3, this vertex
must vanish at $\gamma_{A}=\pi$ since here $H_{2}$ and $H_{3}$ exchange their CP properties they have at $\gamma_{A}=0$.
Moreover, as is clear from a comparison of the two windows, with increasing $\tan\beta$ the entire vertex gets suppressed due to
the fact that 
the decoupling limit is approached \cite{pokorski,decoupling}. The behaviour of the solid curves follows from the CP
compositions of $H_{1}$ and $H_{3}$ in Figs. 1 and 3, and their interpretation proceeds as in $Z H_{1}H_{2}$ vertex. In both 
figures, inclusion of the top quark contribution does not lead to a significant change as its $\tan\beta$ dependence is similar
to the tree vertex. For small $\gamma_{A}$ results of the CP--conserving theory are valid because coupling of the pseudoscalar 
components of $H_{1}$ and $H_{3}$ to identical top squark mass eigenstates is both small, and is a two--loop effect. The latter
statement follows from the fact that the pseudoscalar coupling to two identical top squarks is pure imaginary and it does not
interfere with the tree vertex. This remains true until the light top squark triangle develops an absorbtive part which happens
when the light stop mass falls below $\sqrt{s}/2$. Indeed, the sharp change in the behaviour of the dotted curve
near $\gamma_{A}\sim 1$, which is clear in the upper window, confirms this expectation. For $1\simlt \gamma_{A} \simlt 5$  the
light stop triangle keeps having a non--vanishing absorbtive part because of which the vertex radiative corrections are enhanced.
Here one onserves that $|\hat{V}_{H_{1}H_{3} Z}|$ (dotted curve) is no longer symmetric with respect to $\gamma_{A}=\pi$ because 
of $\sim \sin \gamma_{A} \cos \gamma_{A}$ type dependence of the pseudoscalar coupling to the identical top squarks. For
$\tan\beta=30$, the mass of the light top squark is already below $\sqrt{s}/2$ in the entire range of $\gamma_{A}$ so
that interference effects due to light top squark triangle are at work. However, due to large $\tan\beta$ $|\hat{V}_{H_{1}H_{3}
Z}|$ is suppressed, and vertex radiative corrections remain small compared to the tree vertex.
   
\section{Conclusion}
As described by Figs. 1-3 the Higgs bosons of the MSSM do not have definite CP properties due to explicit CP violation. However,
one notices that the lightest Higgs boson has its CP--odd component much smaller than the CP--even ones, and the heavier Higgs
bosons have alternating CP compositions as $\gamma_{A}$ varies. The pair production process discussed here enables one to probe 
the CP compositions of the Higgs bosons.  In general, from the analysis of $|\hat{V}_{Z H_{1}H_{j\neq 1}}|$ one concludes that
$e^{+}e^{-}\rightarrow H_{1}H_{j\neq 1}$ 
cross section is enhanced or suppressed depending on the value of CP violating phase $\gamma_{A}$. Associated production of the
lightest Higgs with heavier ones is generally suppressed with increasing $\tan\beta$ \cite{decoupling}. This statement has been 
confirmed by Figs. 4 and 5; however, there are certain values of $\gamma_{A}$ for which $e^{+}e^{-}\rightarrow H_{1}H_{3}$
cross section is twice or more larger than its value in the CP--conserving theory.  Needless to say, $e^{+}e^{-}\rightarrow
H_{1}H_{2}$ cross section is completely new as it does not exist in the CP--respecting limit. In this sense, suppression 
of the pair production amplitude is less than the one expected in the CP--conserving MSSM for large $\tan\beta$
\cite{decoupling}. However, we stress that such enhancements follow from the radiative corrections 
in the Higgs sector rather than the vertex radiative corrections. In this sense, it is the CP violation in the Higgs
sector that is important; as already discussed in \cite{pokorski} the vertex radiative corrections are small.
As described here, discussion of the pair production process enables one to have a joint analysis of the
even-- and odd-- CP Higgs bosons in the MSSM. On the other hand, one recalls that the well--known Bjorken process
$e^{+}e^{-}\rightarrow Z H_{j}$ is expected to yield three distinct CP--even Higgs bosons for large enough collider energies if
there is explicit CP violation in the MSSM. Such an unusual signature follows from the violation of the CP symmetry.
Therefore, one concludes that the Higgs search at future
$e^{+}e^{-}$ machines NLC \cite{nlc} and TESLA \cite{tesla} could be highly important for determining the fate of the 
supersymmetric CP violation.

\end{document}